\documentclass[conference,letterpaper]{IEEEtran} 

\usepackage[utf8]{inputenc}
\usepackage[T1]{fontenc}
\usepackage[hidelinks]{hyperref}
\usepackage{url}
\usepackage{ifthen}
\usepackage{cite}
\usepackage[cmex10]{amsmath} 
\usepackage{amsfonts,amssymb}
\usepackage{color}
\usepackage{graphicx}
\usepackage{ifthen}
\usepackage[arxiv]{optional} 
\usepackage{caption}
\usepackage{subcaption}

\def\cX{\mathcal{X}}

\def\cZ{\mathcal{Z}}

\def\cP{\mathcal{P}}

\newcommand{\eq}[1]{\begin{align}#1\end{align}}

\newcommand{\lb}[1]{\left\{ \begin{array}{ll} #1 \end{array} \right.}

\newcommand{\cA}{\mathcal{A}}

\newcommand{\tgamma}{\tilde{\gamma}}

\newcommand{\defeq}{\buildrel\triangle\over =}

\newcommand{\pushright}[1]{\ifmeasuring@ #1 \else\omit\hfill$\displaystyle#1$\fi\ignorespaces}
\newcommand{\pushleft}[1]{\ifmeasuring@ #1 \else\omit$\displaystyle#1$\hfill\fi\ignorespaces}
\newcommand{\nn}{\nonumber}

\newcommand{\E}{\mathbb{E}}

\begin{document}
\title{Social herding in mean field games}
\author{
 \IEEEauthorblockN{Deepanshu Vasal } 
  \IEEEauthorblockA{Northwestern University, \\Evanston, IL\\ \small{{dvasal@umich.edu}}}
 }

\maketitle

\begin{abstract}
In this paper, we consider a mean field model of social behavior where there are an infinite number of players, each of whom observes a type privately that represents her preference, and publicly observes a mean field state of types and actions of the players in the society. The types (and equivalently preferences) of the players are dynamically evolving. Each player is fully rational and forward-looking and makes a decision in each round $t$ to buy a product. She receives a higher utility if the product she bought is aligned with her current preference and if there is a higher fraction of people who bought that product (thus a game of strategic complementarity). We show that for certain parameters when the weight of strategic complementarity is high, players eventually herd towards one of the actions with probability 1 which is when each player buys a product irrespective of her preference. 
\end{abstract}
\section{Introduction}
Understanding social behaviors is an important problem of how agents interact among each other and make decisions in the real world. There are many instances in the real world where people mimic the actions of other players irrespective of their preferences, for instance, people visit the Mona Lisa painting at Louvre because of its fame, visit a famous monument, or buy a product with a lot of positive reviews, etc. Authors in~\cite{Ba92,BiHiWe92} considered a stylized model where there are an infinite number of players who act sequentially once in the system. Each player makes a private observation about the state of the system and observes the actions of the previous players, based on which she makes a decision to either buy or not buy a product. The authors show that players eventually fall into an \emph{informational cascade} with probability 1 where each player discards her private observation and mimics the action of her predecessor. This demonstrates that herding is a rational behavior. Since then there have been a number of papers extending this model or studying other models of Bayesian and non-Bayesian social learning, with or without fully rational players~\cite{SmSo02,AcDaLoOz11,LeSuBe17,MoTa13,MoSlTa14b,MoSlTa15,HaMoStTa14,MoSaJa16,JaMoSaTa12}. For the Bayesian models considered in the current literature including~\cite{BiHiWe92,SmSo02,AcDaLoOz11,LeSuBe14}, each buyer participates only for one time period. To the best of our knowledge, \cite{VaAn22,BiHeAn22,HeBiAn19} are the only other works that consider fully rational forward-looking agents. More specifically~\cite{VaAn22} presents a general framework with finite $N$ agents to study Bayesian learning while \cite{BiHeAn22,HeBiAn19} presents a model with $N$ agents and further extend it to the case when the agents tend to $\infty$. The reason this problem is hard when the players are fully rational and forward-looking is that the corresponding equilibrium concept is perfect Bayesian equilibrium (PBE) and it is very hard to compute~\cite{VaSiAn19,VaAn16cdc} and thus it is not clear if it is even played by players in the real world.

To model the behavior of large population strategic interactions, mean-field games were introduced independently by Huang et al in~\cite{HuMaCa06}, and Larsy and Lions in~\cite{LaLi07}. In such games, there is a large number of homogenous strategic players, where each player has an infinitesimal effect on system dynamics and is affected by other players through a mean-field population state. There have been a number of applications such as economic growth, security in networks, oil production, volatility formation, and population dynamics (see ~\cite{La08,GuLaLi11,SuMa19,HuMa16,HUMa17,HuMa17cdc, AdJoWe15} and references therein). 

In this paper, we consider a model of social interaction based on mean field games where we assume there are an infinite number of players who are fully rational and forward-looking and act throughout the (infinite horizon of the) game. There are two products available and each player in each time $t$ has a binary-valued preference for the products. Each player makes a decision to either buy or not in each round where her instantaneous reward depends on if she bought the product according to her preference and if more people bought that product. Such games are called games of strategic complementarities where a player's utility increases if more people use the same product such as in online gaming, use of language, dating apps, and more. In this paper, we show that in this model, the players herd towards an action (i.e. choose a product irrespective of their preference) with probability 1.  As in~\cite{Ba92,BiHiWe92}, the players are aware that they are herding i.e. actions of the previous players do not reveal any information  about their private preferences. 
This is one of the very few papers in the social herding literature where agents are fully rational and forward-looking.
This work is different from the fully rational forward-looking models considered in\cite{VaAn22, BiHeAn22,HeBiAn19} either in the number of players, reward structure or dynamics. 

The paper is structured as follows. We present the model in Section~II. We present preliminaries on MFE in Section~III. We present the MFE of the game and its asymptotic analysis in Section~IV. We conclude in Section~V. All proofs are presented in the Appendix.
\section{Model}
We assume there are two kinds of technologies A (or `1') and B (or `-1') and there are an infinite number of players where player $i$ observes a binary-valued random variable $x_t\in\{-1,1\} $ privately which determines her preference for the product. Player $i$ takes action $a_t\in\{-1,1\}$ at time $t$ which represents choosing one of the two technologies. The preferences of the players also evolve in an independent Markovian fashion such that
\eq{
P(x_{t+1}\neq x_t|a_t) =  \lb{ p^1 \text{  if  } a_t = x_t\\
 p^2 \text{  if  } a_t \neq x_t,
}
}
where we assume that $p^1<p^2<1/2$. This indicates that there is some sense of ``stickiness" or inertia with the product such that if a follower has a preference for product A and chooses product A, the probability that her preference would change to product B is lower than if she chose the product B in the first place. 
Let $z_t$ be the type mean field state where for $x\in\{-1,1\}$
\eq{
z_t(x) = \lim_{N\to \infty} \frac{1}{N} \sum_{i=1}^N1(x_t = x)
}
and let $\mu_t$ be the action mean-field state of the players where for $a\in\{-1,1\}$,
\eq{
\mu_t(a) = \lim_{N\to \infty} \frac{1}{N} \sum_{i=1}^N1(a_t = a)
}
Each player takes action $a_t \sim \sigma_t(\cdot|z_{1:t},x_{1:t})$.
The mean field state evolves through the Fokker Planck equation
\eq{
z_{t+1}(\cdot) = \sum_{x_t}z_{1:t}(x_{1:t})\sigma(a_t|z_{1:t},x_{1:t})Q(x_{t+1}|x_t,a_t,z_t)
}
Utility for a user depends on her personal preference and is also directly proportional to the number of other users who use that product. Each player gets a reward
\eq{
R(x_t,a_t,\mu_t) = \alpha x_ta_t +(1-\alpha) a_t(2\mu_t(1)-1) 
}
where $\alpha\in[0,1]$ is the weight of personal preference in the reward that a player gets. This reward structure implies that if she chooses the product in accordance with her preference she gets a reward of 1 and -1 otherwise. She also gets a reward proportional to the number of people using that product (thus a game of strategic complementarity). Her final instantaneous reward is a convex combination of these two rewards using weight $\alpha$.

\section{Preliminaries: MFE }

\subsection{Mean field equilibrium (MFE)}
MFE is defined through a forward backward system of equations where the forward equations are Fokker-Planck-Kolmogorov equations and define the evolution of the mean field, while the backward equations are HJB equations and help in designing optimal strategies of the players given a mean field trajectory. The same author in~\cite{Va20acc} presented a sequential decomposition algorithm to find the MFE of MFGs. In this paper, we will use that framework to find the MFE of the game considered. 

As mentioned before, in MFE, strategies of player $i$ depend on the mean-field population state at time $t$, $z_{t}$, and on its current type $x_t$. Equivalently, player $i$ takes action of the form $A_t\sim \sigma_t(\cdot\vert z_t,x_t)$. Similar to the common agent approach in~\cite{NaMaTe13}, an alternate and equivalent way of defining the strategies of the players is as follows. We first generate partial function $\gamma_t:\cX\to\cP(\cA)$ as a function of $z_t$ through an equilibrium generating function $\theta_t:\cZ\to(\cX\to\cP(\cA))$ such that $\gamma_t = \theta_t[z_t]$. Then action $A_t$ is generated by applying this prescription function $\gamma_t$ on player $i$'s current private information $x_t$, i.e. $A_t\sim \gamma_t(\cdot\vert x_t)$. Thus $A_t\sim \sigma_t(\cdot\vert z_{t},x_t) = \theta_t[z_t](\cdot\vert x_t)$.

We are only interested in symmetric Markovian equilibria of such games such that $A_t\sim \gamma_t(\cdot\vert x_t) = \theta_t[z_t](\cdot\vert x_t)$ i.e. strategies are independent of the identities of the players.

For a given symmetric prescription function $\gamma_t = \theta[z_t]$, the statistical mean-field $z_t$ evolves according to the discrete-time Fokker Planck equation~\cite{ArMa14}, $\forall y\in\cX$:
\eq{
z_{t+1}(y) &=\sum_{x\in\cX}\sum_{a\in \cA} z_t(x)\gamma_t(a\vert x)Q_x(y\vert x,a,z_t), \label{eq:z_update}\\
\text{which implies,\;\;\;\;}z_{t+1}&= \phi(z_t,\gamma_t).
}

It is easy to see that
\eq{
\mu_t(a_t) &= \sum_{x_t}z_t(x_t)\gamma_t(a_t|x_t)\\
\text{i.e.\;\;\;\;\;}\mu_t &= G(z_t,\gamma_t)
}
Based on this, we define a backward recursive algorithm to compute MFE as follows.
We define an equilibrium generating function $\theta$, where $\theta:\cZ\to\{\cX\to\mathcal{P}(\cA) \}$, where for each $z$, we generate $\tgamma = \theta[z]$. In addition, we generate a reward-to-go function $V$, where $V:\cZ\times\cX\to\mathbb{R}$.
These quantities are generated through a fixed-point equation as follows.

 $\forall z$, let $\theta[z] $ be generated as follows. Set $\tilde{\gamma} = \theta[z]$, where $\tilde{\gamma}$ is the solution of the following fixed-point equation, $x\in \cX$,
  \eq{
 &\tilde{\gamma}(\cdot\vert x) \in  \arg\max_{\gamma(\cdot\vert x)} \E^{\gamma(\cdot\vert x)} \big[ R(x,A,z) +\nn\\
 &\delta V(\phi(z,\tgamma), X^{'}) \vert  z,x\big] , \label{eq:m_FP}\\
&V(z,x) \defeq \nn\\
& \E^{\tilde{\gamma}(\cdot\vert x)} \big[ R(x,A,z) +\delta V(\phi(z,\tgamma), X^{'}) \vert  z,x\big].  \label{eq:Vdef}
}
 where expectation in \eqref{eq:m_FP} is with respect to random variable $(A,X^{'})$ through the measure
$\gamma(a\vert x)Q_x(x^{'}\vert x,a,z)$.

Then, an equilibrium strategy is defined as 
\eq{
\tilde{\sigma}_t(a_t\vert z_{1:t},x_{1:t}) = \tilde{\gamma}_t(a_t\vert x_t), \label{eq:sigma_fh}
} 
where $\tilde{\gamma}_t = \theta[z_t]$. 

\section{MFE of the game}
We define herding as when action $a_t$ is independent of $x_t$ i.e. when the equilibrium strategies of the players do not depend on their private information.
Numerically we observe a threshold phenomenon at $\alpha= 0.16$ where for $\alpha\leq 0.16$ players take actions irrespective of their private information and action mean field converges to 0 while for $\alpha>0.58$ herding does not occur i.e. actions do depend on players private information. For $0.16<alpha\leq0.58$ herding may occur depending on the initial mean field state. 

We assume $z_0=0.5$. For $\delta =0.9,p^1=0.1,p^2=0.3,\alpha = 0.1, \forall z, \theta[z](a = 1|x_t)= 0$ for $x_t = -1,1,$ and $V(z,-1)=9.53,V(z,1)=9.1 $ satisfy~\eqref{eq:m_FP}. 

\eq{
&V(z,x=-1)= \nn\\
 & \lb{0.1 +0.9  +0.9((1-0.1)\times 9.53+ 0.1\times 9.1) 
  \text { for } a_t = -1\\
  -0.1 -0.9 +0.9((1-0.3)\times 9.53+ 0.3\times 9.1)
  \text { for } a_t = 1
}
}

\eq{
&V(z,x=1)=\nn\\
&\lb{ -0.1 +0.9 +0.9(0.3\times 9.53+ (1-0.3)\times 9.1)
\text { for } a_t = -1\\
0.1 -0.9 +0.9 (0.1\times 9.53+ (1-0.1)\times 9.1)
\text { for } a_t = 1
}
}
Moreover $\mu_t(1)= 0$ and $z_t(1) \to 0.25$. Thus for $\alpha=0.1$ herding occurs with probability 1. 

For $\alpha = 0.9$ herding does not occur. 
For $\delta =0.9,p^1=0.1,p^2=0.3,\alpha = 0.9, \forall z, \theta[z](1|x=-1) = 0,\theta[z](1|x=1) = 1, $ and $V(z,-1)=9.18 - 0.47 z(1),V(z,1)=8.71 + 0.74z(1) $ satisfy~\eqref{eq:m_FP}. Here $\mu_t(1) \to 0.5$ and $z(1)\to 0.5$.

We also plot the utility and equilibrium strategies for $\alpha = 0.2$ in Figures 1-3. We numerically observe that if the initial mean field state is close to 0 or close to 1, players do herd to 0 or 1 respectively, however, if the mean field initial state is in the middle then herding doesn't occur as the equilibrium strategies of the players depend on their private information. For instance, for $\alpha = 0.2, \delta =0.9,p^1=0.1,p^2=0.3$,
\eq{
\theta[z](a=1|x=-1) = \lb{0 \text{ if } z(1)\leq 0.33\\1\text{ if } z(1)>0.33 }
},
\eq{
\theta[z](a=1|x=1) = \lb{0 \text{ if }  z(1)\leq 0.66\\1 \text{ if } z(1)>0.66  }
}
and 
\eq{
V(z,-1)&=\lb{8.8 \text{ if } z(1)\leq 0.33\\
-3.3z(1) + 4.67\text{ if } 0.33<z(1)\leq0.66\\
7.5\text{ if } z(1)>0.66 },\\
V(z,1)&=\lb{7.5 \text{ if } z(1)\leq 0.33\\
3.3z(1) + 1.23\text{ if } z(1)>0.33\\
8.8\text{ if } z(1)>0.66 } 
}
satisfy~\eqref{eq:m_FP}. Here 
\eq{
\mu_t(1) \to \lb{
0 \text{ if } \mu_0\leq 0.33\\
0.5\text{ if } 0.33<\mu_0<0.66\\
1\text{ if } 0.66<\mu_0\leq 1 
}
}
and
\eq{
z(1)\to\lb{0.33 \text{ if } \mu_0\leq 0.33\\
0.5\text{ if } 0.33<\mu_0<0.66\\
0.66\text{ if } 0.66<\mu_0\leq 1 }
}

\begin{figure}
         \centering
         \includegraphics[width=5cm]{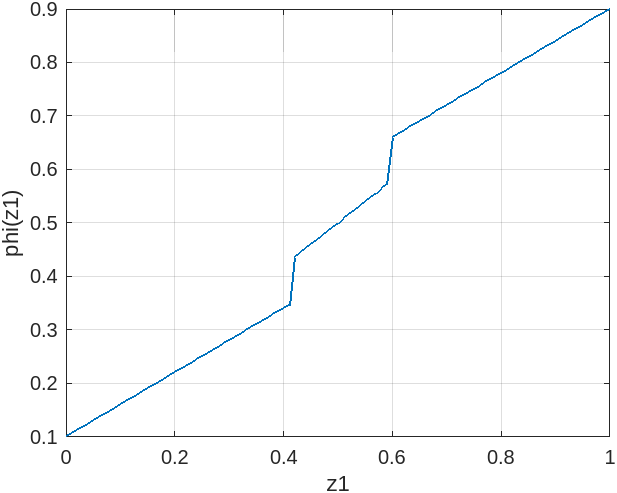}
         \caption{ $\phi(z(1),\tgamma)$}
         \label{fig:y equals x}
     \end{figure}
     \begin{figure}
         \centering
         \includegraphics[width=5cm]{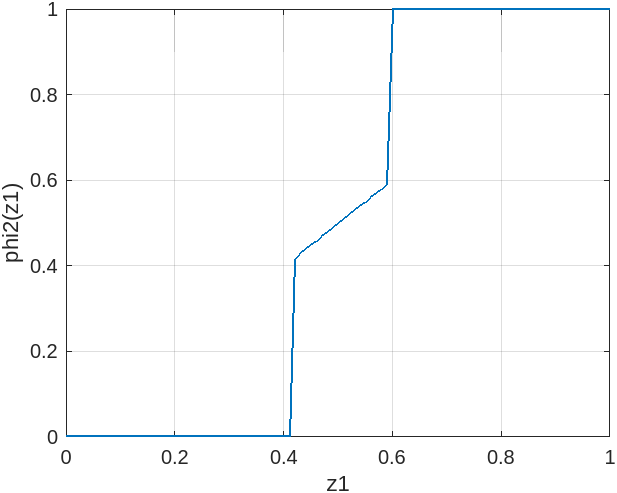}
         \caption{$G(z(1),\tgamma)$}
         \label{fig:three sin x}
     \end{figure}

     \begin{figure}
         \centering
         \includegraphics[width=5cm]{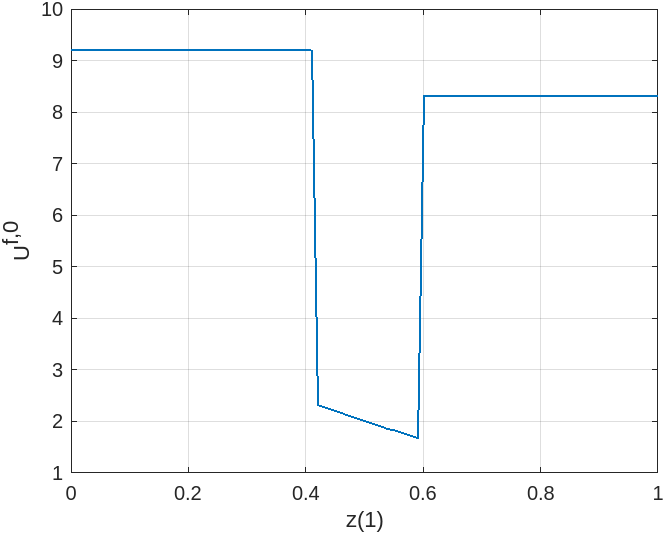}
         \caption{Utility of the user when the state is low}
         \label{fig:y equals x}
     \end{figure}
     \begin{figure}
         \centering
         \includegraphics[width=5cm]{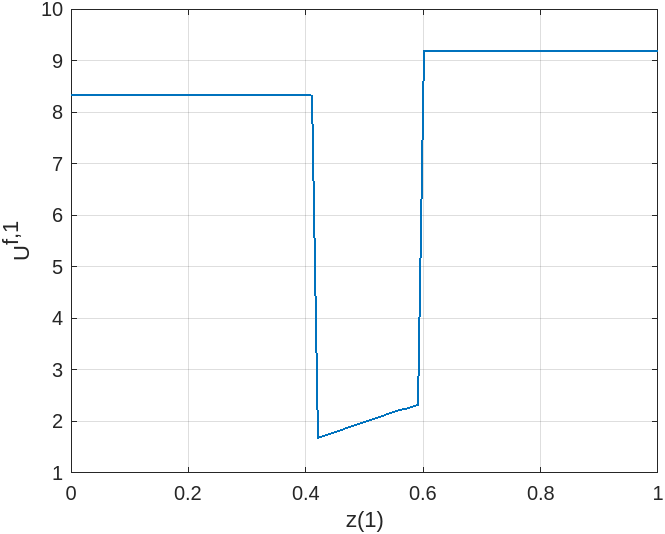}
         \caption{Utility of the user when the state is high}
         \label{fig:three sin x}
     \end{figure}

     \begin{figure}
         \centering
         \includegraphics[width=5cm]{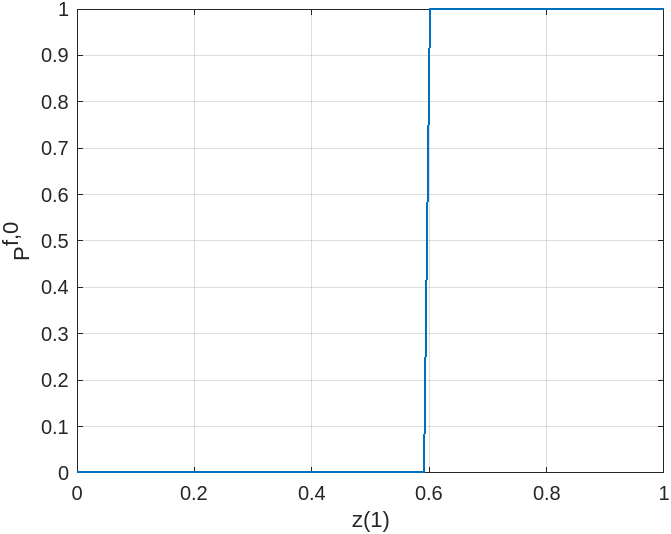}
         \caption{Probability of taking action 1 when the state is low}
         \label{fig:y equals x}
     \end{figure}
     \begin{figure}
         \centering
         \includegraphics[width=5cm]{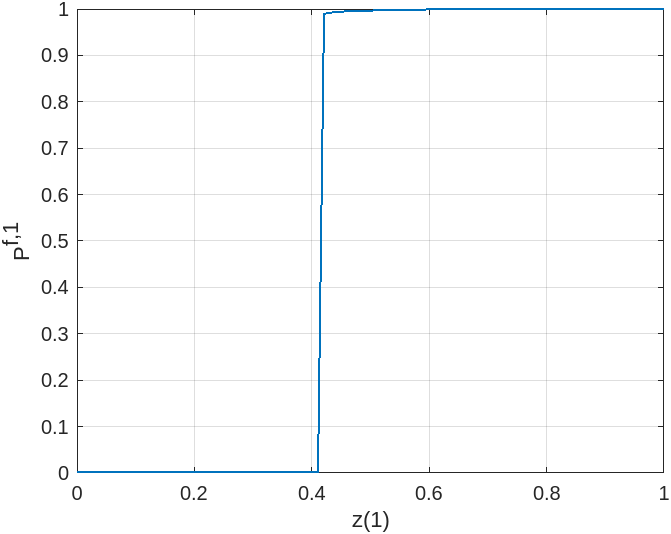}
         \caption{Probability of taking action 1 when the state is high}
         \label{fig:three sin x}
     \end{figure}

This shows how agent's reward structures can affect herding behavior in a society with long range rational agents. More specifically, for stronger network effects, herding occurs almost surely and for weaker network affects it doesn't.

\section{Conclusion}
In this paper, we consider a discrete-time social interaction model based on mean field games where there are an infinite number of players each with a type that denotes their preference that is dynamically evolving. Each player in each time $t$ makes a decision to buy one of the two available products and receives an instantaneous reward that is higher if the player buys a product that is aligned with her preference and if there is a higher fraction of people using that product. We compute its mean field equilibrium and show that players eventually herd almost surely for certain parameters of the problem whereas for certain other parameters they don't, where herding is defined as the state when from then on players' actions don't depend on their private information anymore. Our analysis highlights how reward structures (with strategic complentarity) can play a significant role in herding behavior in the society. This is one of the very few examples in the literature where one can demonstrate the occurrence of herding with fully rational, forward-looking agents and more generally shows how mean field games can be an excellent framework for studying social interactions including herding with fully rational agents. 
 \bibliographystyle{IEEEtran}
 \bibliography{deepanshu,library}
\end{document}